\documentclass[iop]{emulateapj}

\begin{document}

\title{Wind-shearing in gaseous protoplanetary disks \\and the evolution of
binary planetesimals}

\author{Hagai B. Perets and Ruth A. Murray-Clay}

\affil{Harvard-Smithsonian Center for Astrophysics, 60 Garden st. Cambridge
MA 02338, USA }
\begin{abstract}
One of the first stages of planet formation is the growth of small
planetesimals and their accumulation into large planetesimals and
planetary embryos. This early stage occurs much before the dispersal
of most of the gas from the protoplanetary disk. Due to their different
aerodynamic properties, planetesimals of different sizes and shapes
experience different drag forces from the gas during this time. Such
differential forces produce a wind-shearing (WISH) effect between
close by, different size planetesimals. For any two planetesimals,
a WISH radius can be considered, at which the differential acceleration
due to the wind becomes greater than the mutual gravitational pull
between the planetesimals. We find that the WISH radius could be much
smaller than the \emph{gravitational} shearing radius by the star
(the Hill radius). In other words, during the gas-phase of the disk,
WISH could play a more important role than tidal perturbations by
the star. Here we study the WISH radii for planetesimal pairs of different
sizes and compare the effects of wind and gravitational shearing (drag
force vs. gravitational tidal force). We then discuss the role of
WISH for the stability and survival of binary planetesimals. Binaries
are sheared apart by the wind if they are wider than their WISH
radius.  WISH-stable binaries can also inspiral, and possibly coalesce, due to gas drag. 
Here, we calculate the WISH radius and the gas drag-induced merger timescale, providing
stability
and survival criteria for gas-embedded binary planetesimals. Our results
suggest that even WISH-stable binaries may merge in times shorter
than the lifetime of the gaseous disk. This may constrain currently
observed binary planetesimals to have formed far from the star or
at a late stage after the dispersal of most of the disk gas. We note
that the WISH radius may also be important for other processes such
as planetesimal erosion and planetesimal encounters and collisions
in a gaseous environment. 
\end{abstract}

\section{Introduction}

Gravitational encounters between planetesimals play an important role
in the evolution of protoplanetary disks and planet formation \citep[e.g., reviews by][]{lis+93,gol+04}.
Planetesimal growth likely occurs while the planetesimals are still
embedded in a gaseous disk. Studies of gas-planetesimal interactions
have shown that gas can affect the velocity dispersion of planetesimals
\citep[e.g.][]{nel+10}, may help the formation of large planetesimals
through clumping of planetesimals \citep[][and references therein]{chi+10},
and can lead to fast inspiral of planetesimals into the star through
gas-drag \citep[][and references therein]{nak+86,wei77}. Here we focus
on a different aspect of planetesimals embedded in a gaseous disk,
namely the close interaction between pairs of single planetesimals
in a gaseous environment.

Planetesimals likely vary in size and shape, and therefore have a
wide range of aerodynamical properties, which affect their interaction
with surrounding gas. In particular, planetesimals of different sizes
and/or shapes experience different drag forces from the head wind
they encounter in the gaseous disk. The difference between the forces
acting on two different-size planetesimals \citep{wei77} can change
their relative trajectories with respect to their unperturbed motion
in the absence of gas (for example \citealp{orm+10} considered planetesimal
interactions in gas rich environment; their study focused on planar
encounters and drag law regimes which are linearly dependent on velocity).

During an encounter between two different-size planetesimals, the
different forces experienced by the two components as a result of
gas drag generate a wind-shearing (WISH) effect, which could be stronger
than their gravitational interaction. For any two planetesimals, we
consider the radius, which we term the WISH radius, at which the differential
acceleration due to aerodynamical wind-shearing becomes greater than
the mutual gravitational pull between them. In the following we explore
this new distance scale and discuss its implications, including the
stability of binary planetesimals. In addition, we study the evolution
of WISH-stable binary planetesimals in gas. Such binaries dissipate their orbital
energy through gas drag and may inspiral to form closer binaries or
even coalesce during the typical lifetime of a protoplanetary disk.

We begin by deriving the WISH radius and discussing the effects of
gas drag on particles of different sizes. We then calculate the WISH
radius for two planetesimals of arbitrary effective sizes (Section
\ref{sec:radcalc}). In Section \ref{sec-evol}, we consider the evolution of binary planetesimals 
embedded in a gas disk, including WISH stability (Section \ref{sec:binstab}) and gas
drag-induced inspiral (Section \ref{sec-inspiral}). 
Finally, we discuss various other possible implications
of our results (Section \ref{sec:other}) and summarize (Section \ref{sec:sum}).

\pagebreak

\section{Gas drag and the wind-shearing radius in protoplanetary disks}

\label{sec:radcalc}

Planetesimals of different sizes embedded in the same gaseous environment
experience different drag forces and hence different accelerations.
The differential acceleration between two planetesimals of mass $m_{b}$
and $m_{s}$ due to the wind-shearing effect is given by \begin{equation}
\Delta a_{WS}=\left|\frac{F_{D}(m_{b})}{m_{b}}-\frac{F_{D}(m_{s})}{m_{s}}\right|=\frac{3\rho_{p}}{4\pi}\left|\frac{F_{D}(r_{b})}{r_{b}^{3}}-\frac{F_{D}(r_{s})}{r_{s}^{3}}\right|,\;\label{eq:differential-acceleration-general}\end{equation}
 where $F_{D}$ is the force exerted on a particle due to gas drag.
Throughout this paper, we perform our calculations for spherical particles
of constant density, $\rho_{p}$, so that the mass of a planetesimal
with radius $r$ is $m=(4/3)\pi\rho_{p}r^{3}$. In Equation (\ref{eq:differential-acceleration-general}),
planetesimal masses $m_{b}$ and $m_{s}$ correspond to radii $r_{b}$
and $r_{s}$, respectively. Real planetesimals could have different
aerodynamical properties (e.g. they may not be spherical and/or they
could be porous); however, calculations analogous to those presented
here may be performed for any form of $F_{D}(m)$.

For small separations over which the environmental conditions (gas
density and temperature) are approximately the same, the differential
WISH acceleration between any two planetesimals is independent of
the distance between them. In this case, $\Delta a_{WS}$ can be used
to define an important distance scale, which we term the WISH radius.
To define this scale, we adopt a similar approach to that used to
define the the gravitational tidal-shearing radius, i.e., the Hill
radius.

The Hill radius (sphere) is the distance at which the gravitational
influence of a planetesimal or a planet with mass $m$ and radius
$r$, orbiting a star with mass $M_{\star}$ at radial distance $a$,
becomes comparable to the tidal perturbation by the star. It is given
by \begin{equation}
R_{H}=\left(\frac{m}{3(M_{\star}+m)}\right)^{1/3}a\simeq\left(\frac{4\pi\rho_{p}}{9M_{\star}}\right)^{1/3}ra,\label{eq:R_H}\end{equation}
 where the second expression is derived for a spherical planetesimal
with $m\ll M_{\star}$. A test particle located close to the Hill
radius, or beyond it, is strongly affected by the gravitational pull
of the star. If it begins in orbit around the planetesimal, its orbit
will be perturbed and is likely to become unstable. The exact distance
up to which a binary orbit can remain stable also depends on its orbit
direction, e.g. prograde or retrograde with respect to the orbit of
the planet around the star \citep{ham+91,she+08,per+09}, or more
generally, the relative inclination of the particle's orbit. In the
following we adopt the simple definition given by Eq. (\ref{eq:R_H}).

Following the definition of the Hill radius we can now define the
WISH radius. This radius is defined as the distance between two planetesimals
for which the differential WISH acceleration between them equals their
mutual gravitational pull. Equating $\Delta a_{WS}$ with the gravitational
acceleration $a_{grav}=G(m_{b}+m_{s})/d_{bin}^{2}$ yields a separation
$d_{bin}$ between the two planetesimals equal to the WISH radius,
which we define as \begin{equation}
R_{WS}=\sqrt{\frac{G(m_{b}+m_{s})}{\Delta a_{WS}}}.\label{eq:R_WS}\end{equation}
 Beyond this limiting radius even two planetesimals which are formally
gravitationally bound (in the absence of WISH) would be sheared apart
by the wind.

In order to calculate the specific value of the WISH radius for any
given pair of planetesimals, we first need to understand the gas-drag
force applied on planetesimals which face a head wind. This depends
on the specific regime of the gas-planetesimal interaction, since
different gas-drag laws apply under different conditions. We review
gas drag laws in Section \ref{sub:Drag-laws}, calculate $\Delta a_{WS}$
explicitly for several regimes in Section \ref{sub:The-aerodynamic-shearing-acceleration},
and combine these to provide self-consistent calculations of $R_{WS}$
as a function of planetesimal size in a fiducial disk (Section \ref{sub:RWS}).

\subsection{Drag laws\label{sub:Drag-laws}}

The appropriate gas-drag force on a planetesimal of radius $r$ moving
through gas at relative velocity $v_{rel}$ depends on $r/\lambda$,
where $\lambda=\mu/(\rho_{g}\sigma)$ is the mean free path of the
gas, $\sigma$ is the cross-section for gas-gas collisions, and $\mu$
is the mean molecular weight. For planetesimals with diameters larger
than the mean free path of the gas, the drag force also depends on
the fluid Reynolds number $Re=2rv_{rel}/(0.5\bar{v}_{th}\lambda)$.
Here, $2r$ is the diameter of the planetesimal. The gas has kinematic
viscosity $(1/2)\bar{v}_{th}\lambda$, temperature $T$, and mean
thermal velocity (for a Maxwellian distribution) $\bar{v}_{th}=(8/\pi)^{1/2}c_{s}$,
where $c_{s}=(kT/\mu)^{1/2}$ is the sound speed and $k$ is Boltzmann's
constant. The various gas drag regimes can be summarized as follows
(where we follow \citealp{wei77}, who in turn follows \citealp{whi73}).\footnote{Planetesimals
move through the protoplanetary disk at subsonic velocities.  For $v_{rel} > c_s$, ram pressure drag applies (c.f. Equation \ref{eq:drag-laws-big}).}

When $r\lesssim\lambda$, drag may be modeled by considering individual
and independent particle collisions, and the Epstein regime applies
(for subsonic $v_{rel}$, which is appropriate for our problem): \begin{equation}
F_{D}=\frac{4}{3}\pi\rho_{g}\bar{v}_{th}v_{rel}r^{2},\label{eq:Epstein-drag}\end{equation}
 For $r\gtrsim\lambda$, the gas must be modeled as a fluid. We take
$r=(9/4)\lambda$ as the boundary between these regimes. At low Reynolds
number, the gas/particle boundary layer dominates (Stokes drag), while
at high $Re$, the gas exerts a Ram pressure force on the particle,
so that the drag law is \begin{eqnarray}
F_{D} & = & 3\pi\rho_{g}\bar{v}_{th}v_{rel}\lambda r\;\;\;\;\;\;\;\;{\rm for}\;{Re}<1\,\,\,\,\,\,\,\,\,\,\,\,\,\,\, Stokes\nonumber \\
F_{D} & = & 0.22\pi\rho_{g}v_{rel}^{2}r^{2}\;\;\;\;\;\;\;\;{\rm for}\;{Re}\gtrsim800\,\,\,\,\,\,\,\,\,\, Ram.\label{eq:drag-laws-big}\end{eqnarray}
 An intermediate regime exists for $1\lesssim Re\lesssim800$. More
generally, the full range of Reynolds numbers can be fitted with a
drag law of \begin{eqnarray}
F_{D} & = & \frac{1}{2}C_{D}(R_{e})\pi r^{2}\rho_{g}v_{rel}^{2}\label{eq:drag-law}\end{eqnarray}
 where $C_{D}(Re)$ can be fitted with an empirical formula based
on recent experimental data in the regime $10^{-3}\le Re\le10^{5}$
(\citealp{bro+03,che09}; compatible with older data used by \citealp{whi73}),
yielding \begin{equation}
C_{D}=\frac{24}{Re}(1+0.27Re)^{0.43}+0.47[1-exp(-0.04Re^{0.38})].\label{eq:Cd}\end{equation}
 We use Equations (\ref{eq:drag-law}) and (\ref{eq:Cd}) for our
drag law in the calculations that follow, except when a single, specific
drag law is specified, in which case we use Equations (\ref{eq:Epstein-drag})
and (\ref{eq:drag-laws-big}).

\subsection{The wind-shearing differential acceleration \label{sub:The-aerodynamic-shearing-acceleration}}

When two particles with radii $r_{b}$ and $r_{s}$ experience gas
drag in the same drag regime, we may obtain a simple expression for
the wind-shearing differential acceleration, $\Delta a_{WS}$. For
example, in the Epstein regime ($r_{s},r_{b}<\lambda$), \begin{equation}
\Delta a_{WS}=\frac{4}{3}\pi\rho_{g}\bar{v}_{th}\left|\frac{r_{b}^{2}v_{rel}(r_{b})}{m_{b}}-\frac{r_{s}^{2}v_{rel}(r_{s})}{m_{s}}\right|\,\,\,\,\,\,\,\,\, Epstein.\label{eq:a_diff_Epstein}\end{equation}
 In general, the relative velocity between each of the planetesimals
and the gas could differ, in which case even planetesimals of the
same size can experience a differential WISH acceleration. For bound
binary planetesimals (which we discuss in Section \ref{sec:binstab})
the velocity relative to the gas of the two components should be,
on average, approximately the same, so that $v_{rel}=v_{rel}(r_{b})=v_{rel}(r_{s})$.
In this case, Eq. (\ref{eq:a_diff_Epstein}) simplifies further: \begin{eqnarray}
\Delta a_{WS} & = & \rho_{g}\bar{v}_{th}v_{rel}\left|\frac{r_{b}^{2}}{r_{b}^{3}\rho_{p}}-\frac{r_{s}^{2}}{r_{s}^{3}\rho_{p}}\right|\nonumber \\
 & = & \frac{\rho_{g}}{\rho_{p}}\frac{\bar{v}_{th}v_{rel}}{r_{s}}\left|\frac{r_{s}}{r_{b}}-1\right|\nonumber \\
 & \simeq & \frac{\rho_{g}}{\rho_{p}}\frac{\bar{v}_{th}v_{rel}}{r_{s}}\;\;,\,\,\,\,\,\,\,\,\, Epstein\label{eq:Epstein-simple}\end{eqnarray}
 where the last expression is for $r_{b}\gg r_{s}$. Similarly, under
the same assumptions, the Stokes regime produces \begin{eqnarray}
\Delta a_{WS} & \simeq & 3\pi\rho_{g}\bar{v}_{th}v_{rel}\lambda\left(\frac{r_{s}}{m_{s}}\right)\nonumber \\
 & = & \frac{9}{4}\frac{\mu}{\rho_{p}\sigma}\frac{\bar{v}_{th}v_{rel}}{r_{s}^{2}},\,\,\,\,\,\,\,\,\, Stokes\label{eq:diffa_stokes}\end{eqnarray}
 and in the Ram pressure regime, \begin{eqnarray}
\Delta a_{WS} & \simeq & 0.22\pi\rho_{g}v_{rel}^{2}\left(\frac{r_{s}^{2}}{m_{s}}\right)\nonumber \\
 & = & 0.165\frac{\rho_{g}}{\rho_{p}}\frac{v_{rel}^{2}}{r_{s}}.\,\,\,\,\,\,\,\,\, Ram\label{eq:diffa_ram}\end{eqnarray}
 We note that in the Stokes regime, $\Delta a_{WS}$ does not depend
on the density of the gas.

More generally, the differential acceleration can be obtained accurately
for any combination of two planetesimals in different drag law regimes
and moving through the gas at different velocities. This can be done
by using Eqs. (\ref{eq:Epstein-drag}) and (\ref{eq:drag-laws-big})
or Eqs. (\ref{eq:drag-law}) and (\ref{eq:Cd}) to calculate the appropriate
drag force on each planetesimal.

\subsection{The wind-shearing radius }

\label{sub:RWS}

Given the expressions in Equations (\ref{eq:Epstein-simple})--(\ref{eq:diffa_ram}),
we can obtain relatively simple formulas for the wind-shearing radius
(Equation \ref{eq:R_WS}) for two planetesimals in the same drag law
regime with $v_{rel}(r_{b})=v_{rel}(r_{s})$ and $r_{b}\gg r_{s}$:
\begin{eqnarray}
R_{WS} & = & \sqrt{\frac{G(m_{b}+m_{s})}{\Delta a_{WS}}}\approx\sqrt{\frac{Gm_{b}}{\Delta a_{WS}}}\label{eq:r_WS_all}\\
 & = & \left(Gm_{b}\rho_{p}\right)^{1/2}\times\left\{ \begin{array}{lr}
{\displaystyle \left(\frac{1}{\rho_{g}\bar{v}_{th}v_{rel}}\right)^{1/2}r_{s}^{1/2}\,} & Epstein,\\
\rule{0ex}{5ex}{\displaystyle \left(\frac{4}{9}\frac{\sigma}{\mu\bar{v}_{th}v_{rel}}\right)^{1/2}r_{s}\,} & Stokes,\\
\rule{0ex}{5ex}{\displaystyle \left(\frac{1}{0.165}\frac{1}{\rho_{g}v_{rel}^{2}}\right)^{1/2}r_{s}^{1/2}\,} & Ram,\end{array}\right.\nonumber \end{eqnarray}
 in the Epstein, Stokes, and Ram pressure regimes, respectively. In
fact, these expressions apply as long as the gas accelerates the smaller
body more effectively than the larger body, even if their drag regimes
are different. In this case the drag regime in Eq. (\ref{eq:r_WS_all})
would correspond to that of the smaller body.

More generally, we can calculate $R_{WS}$ for any two planetesimals
of arbitrary size (in the same or in different gas-drag regimes) as
a function of the properties of the gas in which they are embedded.
When $r_{b}<R_{WS}<R_{H}$, the WISH radius represents the limiting
separation of a binary planetesimal (or a pair of small satellites)
in a gaseous environment.

\begin{figure}
\includegraphics[scale=0.4]{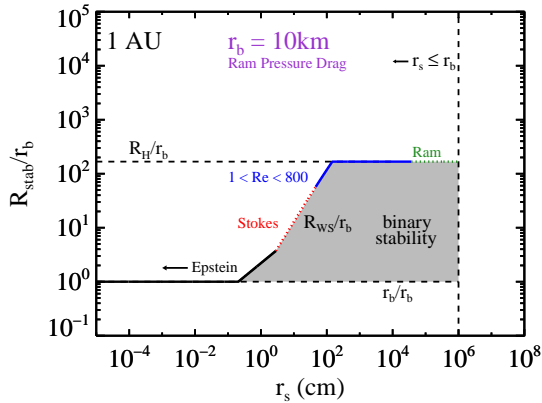}\caption{\label{fig:WISH-stability}The binary stability radius for a $r_{b}=10$
km planetesimal with a companion of radius $r_{s}<r_{b}$, at 1 AU
from the star in our fiducial disk. At radii between the physical
size and the Hill radius of the large planetesimal (lower and upper
dashed lines, respectively), the stability radius equals the WISH
radius, $R_{WS}$ (see text). The planetesimals are moving at a relative
velocity of $v_{rel}\approx0.5c_{s}^{2}/v_{K}$ with respect to the
gas. The relevant drag regime for the small body changes with $r_{s}$,
and the WISH radius is well approximated by Equation (\ref{eq:r_WS_all}).
A pair of bound planetesimals (a binary) can only exist in the shaded
region; it can not reside lower than the physical size (collision),
above the Hill radius (gravitationally unbound by tidal shearing from
the star) or to the left of the WISH radius (sheared apart by the
wind). }

\end{figure}

In Figure \ref{fig:WISH-stability}, we show the calculated $R_{WS}$
and the resulting binary stability radius for a planetesimal with
radius $r_{b}=10$~km, orbited by a smaller body with a range of
sizes. This calculation is performed at $1$ AU from a solar-mass
star in a disk having the following parameters. We choose a disk temperature
of $T=T_{0}(a/{\rm AU})^{-3/7}$ with $T_{0}=120$~K, following \citet{chi+10},
who adapt the results of \citet{chi+97} for a disk around the young
Sun. Varying the value of $T_{0}$ within a reasonable range for Sun-like
stars does not qualitatively change our results. We take the surface
density of the disk to be $\Sigma_{g}=\Sigma_{0}(a/{\rm AU})^{-1}$,
with $\Sigma_{0}=2\times10^{3}$ g/cm$^{2}$. This choice is roughly
consistent with the minimum-mass solar nebula at $1$ AU as well as
with observed dust surface density profiles at distances larger than
$\sim20$ AU in extrasolar disks, taking a dust to gas mass ratio
of $1:100$ (typically assumed in the modeling of protoplanetary disks,
e.g. \citealp{and+10}). Protoplanetary disks likely exhibit a range
of surface density profiles across different systems and at different
times within the same system. We discuss the impact of varying the
gas surface density in the Appendix.

Given the above choices, the disk scale height $H$ is given by \begin{equation}
\frac{H}{a}\sim\frac{c_{s}}{\Omega a}\sim0.022\left(\frac{a}{AU}\right)^{2/7},\end{equation}
 where $\Omega=(GM_{\star}/a^{3})^{1/2}$ is the Keplerian orbital
frequency. The gas volume density profile is then \begin{equation}
\rho_{g}\sim\frac{\Sigma_{g}}{2H}\sim3\times10^{-9}\left(\frac{a}{AU}\right)^{-16/7}\;{\rm g/cm}^{3}.\end{equation}

Using the neutral collision cross-section $\sigma\sim(3$\AA{}$)^{2}\sim10^{-15}$
cm$^{2}$, the mean free path of the gas is \begin{equation}
\lambda\sim\frac{1}{n_{g}\sigma}\sim1\left(\frac{a}{AU}\right)^{16/7}\;{\rm cm}\end{equation}
 where the gas number density $n_{g}=\rho_{g}/\mu$ and we have used
$\mu=2.3m_{H}$ with $m_{H}$ equal to the mass of a hydrogen atom.

For these calculations, we assume that the relative velocity between
the binary and the gas is equal to the velocity of a single planetesimal
with radius equal to that of the larger component of the binary, $r_{b}$,
as it moves through a smooth disk under the influence of gas drag.
This approximation is valid for $r_{b}\gg r_{s}$. In reality, a bound
binary will move through the gas at a velocity that reflects the drag
on both binary components. Following \citeauthor{you10} (2010; see
also \citealp{nak+86}), we set the relative velocity between a planetesimal
and the gas to be $v_{rel}=(v_{rel,r}^{2}+v_{rel,\phi}^{2})^{1/2}$
with \begin{eqnarray}
v_{rel,r} & = & -2\eta v_{K}\left[\frac{t_{s}\Omega}{1+(t_{s}\Omega)^{2}}\right]\label{eqn-vrelr}\\
v_{rel,\phi} & = & -\eta v_{K}\left[\frac{1}{1+(t_{s}\Omega)^{2}}-1\right]\label{eqn-vrelphi}\end{eqnarray}
 with $\eta\equiv(v_{K}-v_{g,\phi})/v_{K}$, so that $\eta v_{K}$
equals the difference between the azimuthal gas velocity, $v_{g,\phi}$,
and the Keplerian velocity, $v_{K}=\Omega a$. We use the approximate
value $\eta=0.5(c_{s}^{2}/v_{K}^{2})$. We calculate the stopping
time $t_{s}=mv_{rel}/F_{D}$ and relative velocity $v_{rel}$ of a
planetesimal iteratively, using the drag law in Equation (\ref{eq:Cd}),
in order to achieve self-consistent values for these and hence for
$F_{D}$ in all drag regimes. Note, however, that our choice of $v_{rel}$
represents the velocity of a planetesimal moving through a uniform
disk and does not take into account turbulence. In addition, even
in a smooth disk planetesimal growth likely occurs in regions of enhanced
solids, which may accelerate the disk gas to more nearly Keplerian
speeds, reducing this relative velocity. We discuss how our results
vary as a function of relative velocity in the Appendix.

Under our assumed conditions, a $10$ km planetesimal at $1$ AU orbits
at approximately the Keplerian velocity, so that in Figure \ref{fig:WISH-stability},
$v_{rel}\approx0.5c_{s}^{2}/v_{K}$. At this relative velocity, the
Reynolds number for a planetesimal with radius $r_{s}$ is \begin{eqnarray}
{\rm Re} & = & \frac{2r_{s}v_{rel}}{0.5\bar{v}_{th}\lambda}\sim\sqrt{\frac{\pi}{2}}\frac{r_{s}}{\lambda}\frac{c_{s}}{v_{K}}\sim\left(\frac{\lambda}{r_{s}}\right)^{-1}\left(\frac{H}{a}\right)\nonumber \\
 &  & \sim0.02\left(\frac{r_{s}}{{\rm 1\; cm}}\right)\left(\frac{a}{AU}\right)^{-2}\end{eqnarray}
 The transition from the Epstein to the Stokes regime for the small
planetesimals may be clearly seen in Figure \ref{fig:WISH-stability}
as a change in the slope of the WISH radius from $1/2$ to unity.
This behavior is matched by Eq. (\ref{eq:r_WS_all}). Though the large
body in this plot always remains in the Ram pressure drag regime,
it is not accelerated much by the gas, and $\Delta a_{WS}$ is dominated
by the acceleration of the small companion. The agreement between
these results and Eq. (\ref{eq:r_WS_all}) reflects the fact that
in this regime, the small companions are accelerated more effectively
by the gas than the $10$ km large body.

\begin{figure}
\includegraphics[scale=0.4]{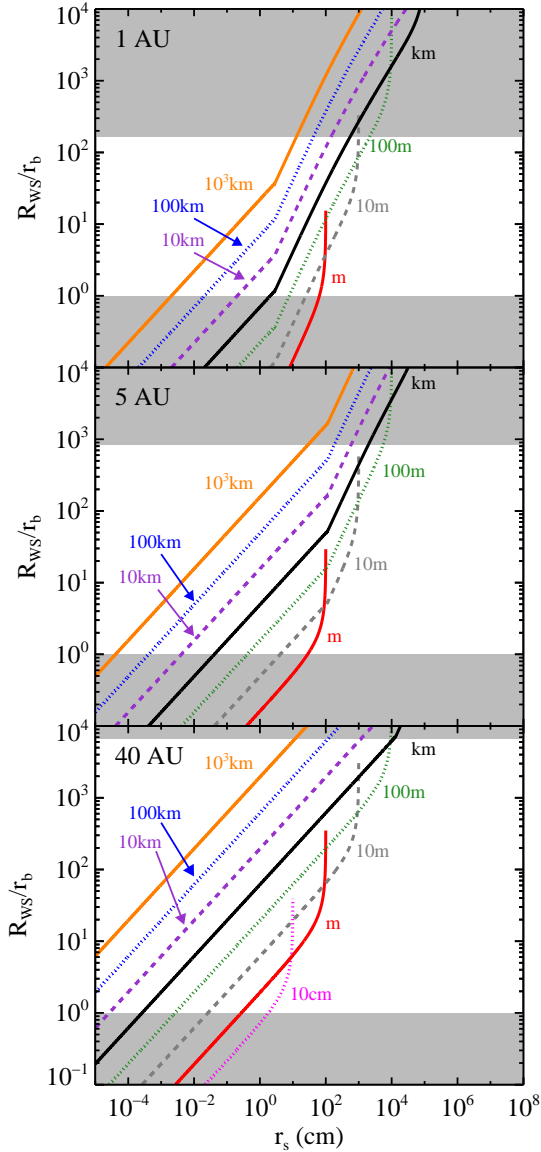}\caption{\label{fig:WISH-radius}The WISH radius, $R_{WS}$, (see text and
Figure \ref{fig:WISH-stability}) for two planetesimals of sizes $r_{s}$
and $r_{b}$ as a function of $r_{s}$ , at distances of 1, 5 and
40 AU from the star (top, middle, and lower panels, respectively).
The plotted lines show $R_{WS}$ for planetesimals with radii $r_{b}=10^{-1}-10^{8}$
cm, in logarithmic jumps. The shaded regimes show the regions where
the WISH radius is smaller than the physical radius or larger than
the Hill radius of the big planetesimal. We do not show lines corresponding
to big planetesimals for which the WISH radius is smaller than the
physical radius unless the smaller planetesimal nearly equals the
larger planetesimal in size (e.g. $r_{b}<1$ m at $1$ and $5$ AU). }

\end{figure}

Figure \ref{fig:WISH-radius} displays the calculated WISH radius as
a function of the small planetesimal size for various sizes of the
large planetesimal and at different distances from the star. Also
shown for comparison are the physical size of the big planetesimal
and its Hill radius. The WISH radius diverges for equal size planetesimals
(with the same velocities relative to the gas), since they experience
the same gas drag, and their differential WISH acceleration approaches
zero.

\section{Binary planetesimals in a gaseous environment}\label{sec-evol}

A non-negligible fraction of currently observed planetesimals in the
Solar system (including asteroids and Trans-Neptunian objects) are
found to be members of binaries \citep[e.g.,][]{ric+06,nol+08b}. Binary
planetesimals can teach us about the dynamical evolution of the Solar
system \citep{per+09,mur+10,par+10} and can play a role in planet
formation and planetesimal growth \citep{nes+10,per10}. Study of
the interactions of binary planetesimals with gas is therefore important
for understanding the formation, stability and evolution of these
binaries and their implications. In the following we discuss the effect
of WISH and gas drag inspiral and coalescence of binary planetesimals
in gas.

\subsection{Wind-shearing disruption of binary planetesimals}

\label{sec:binstab}

We have already alluded to an immediate consequence of the wind-shearing
radius for binary planetesimals, namely that it provides a new stability
criterion for their survival (see Figure \ref{fig:WISH-stability}).
In a gas free environment, binary planetesimals are stable as long
as their separation is smaller than the Hill radius, whereas wider
binaries are destabilized and disrupted by the tidal gravitational
shearing from the star. However, in the presence of gas, the Hill
radius stability limit should be replaced by the WISH radius when
$R_{WS}<R_{H}$ (binaries wider than the Hill radius are always unstable).
We find that the stability criterion for binaries embedded in gas
is \begin{equation}
d_{bin}\le min(R_{H},R_{WS}).\label{eq:stability-criteria}\end{equation}
 Because collisions prevent binaries from forming with $d_{bin}<r_{b}$,
no stable binaries are possible when $R_{WS}<r_{b}$.

\begin{figure}
\includegraphics[scale=0.4]{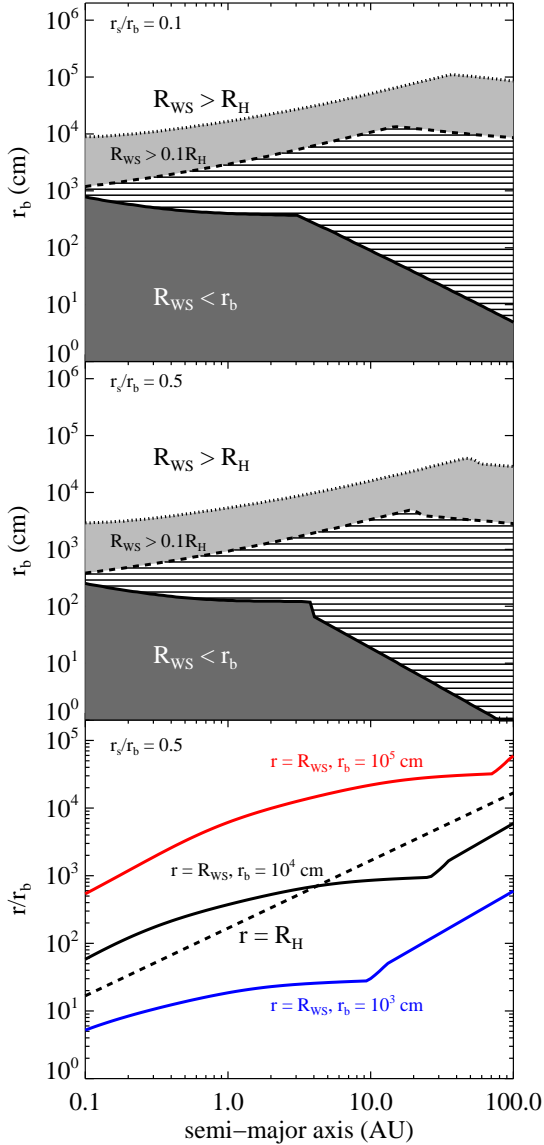}\caption{\label{fig:WISH-sma-dependence} Comparison of the WISH radius and
the Hill radius as a function of planetesimal size and distance from
the star. The WISH radius restricts binary stability most at larger
separations, where the tidal force from the star becomes weak. Top
panel: The planetesimal size for which $R_{WS}(r_{b})=R_{H}(r_{b})$
(dotted line), $R_{WS}=0.1R_{H}$ (dashed line) and $R_{WS}(r_{b})=r_{b}$
(solid line), as a function of the distance from the star, for a binary
planetesimal with size ratio $r_{s}/r_{b}=0.1$. Middle panel: Same
a top panel, but for $r_{s}/r_{b}=0.5$. Larger size ratios reduce
the impact of WISH, making binaries more stable. Binaries in the bottom
shaded region can not survive WISH and will be disrupted by the wind.
The stability of binaries in the top region (above the $R_{W}=R_{H}$
line) is set by the Hill radius rather than by the WISH radius, i.e.
in that region gravitational shearing is stronger than WISH. Bottom
panel: The binary WISH stability radius as a function of the distance
from the star for three binaries with a size ratio $r_{s}/r_{b}=0.5$,
and primaries of $r_{b}=1$ km (top red solid line), $r_{b}=0.1$
km (middle black solid line) and $r_{b}=0.01$ km (bottom blue solid
line). Also shown (dashed line) is the spatial linear dependence of
the Hill radius, corresponding to the stability radius for a non-gaseous
planetesimal disk. Note that $R_{H}/r_{b}$ is independent of $r_{b}$.}

\end{figure}

Given these considerations, the limiting separations of binary planetesimals
in our fiducial disk as a function of size and distance from the star
may be read from Figure~\ref{fig:WISH-radius} (see also Figure \ref{fig:WISH-stability}).
For planetesimal sizes spanning a wide range, $r_{b}<R_{WS}<R_{H}$
in our fiducial disk and this limiting separation is equal to the
WISH radius. Binary planetesimals can therefore be strongly affected
by WISH, most notably for smaller planetesimals closest to the star.
Generally we find that for a wide range of binary and disk properties
the WISH radius determines the stability rather than the Hill radius.
A gaseous environment qualitatively changes the spatial dependence
of binary stability in a protoplanetary disk, as the spatial dependence
(distance from the star) of the Hill radius and that of the WISH radius
differ. This can be seen in Figure \ref{fig:WISH-sma-dependence}, which
compares the WISH radius with the Hill radius for a given binary pair
as a function of the distance from the star (also compare the panels
in Figure \ref{fig:WISH-radius}). We note that binaries with radii
of $\sim100$ km and components of roughly equal mass, comparable
to many observed Trans-Neptunian and asteroid binaries, have their
stability determined by the Hill radius at all distances from the
star in our fiducial disk.

More details regarding the dependence of our results on the gas
density and the relative velocity of planetesimals with respect to
the gas can be found in the Appendix. 

\subsection{Gas drag-induced inspiral of binary planetesimals}\label{sec-inspiral}

Many studies have demonstrated that single planetesimals inspiral
toward their host star due to gas drag (e.g., \citealt{wei77}; for a review see \citealt{chi+10} and references therein).
A similar process can cause the inspiral of a binary planetesimal
into a closer mutual orbit, possibly ultimately leading to coalescence.
In the following we explore the evolution of the mutual orbits of
binary planetesimals in gas, and we provide the timescales for their
coalescence.

For simplicity we restrict our discussion to binary planetesimals
with $m_{s}\ll m_{b}$, but our results can be simply generalized
to arbitrary mass ratios. We study the evolution of a binary orbiting
the Sun in the plane of the protoplanetary disk. For simplicity, we
assume throughout that the mutual binary orbit begins circular and
dissipates orbital energy on a timescale much longer than the mutual
orbital period. The two components of the binary therefore orbit one
another on roughly circular trajectories at a (shrinking) binary separation
of $d_{bin}$. In some circumstances binaries can evolve on faster
timescales; such short term evolution relates to binary planetesimal
formation through gas dissipation and may result in fast coagulation
of planetesimals. These latter processes will be discussed elsewhere
(see Murray-Clay \& Perets, in preparation).

In this limit, the binary loses angular momentum $L$ on a timescale
of $|L/\dot{L}|=m_{s}v_{bin}/\left<F_{D}\right>$, where $v_{bin}$
is the binary orbital velocity of the small body and $\left<F_{D}\right>$
is the gas drag force on the small body, averaged over a binary orbital
period. Equivalently, the binary loses orbital energy $E$ on a timescale
of $|E/\dot{E}|=0.5m_{s}v_{bin}^{2}/(\left<F_{D}\right>v_{bin})$.
Inspiral therefore proceeds on a timescale $\tau_{{\rm merge}}\equiv d_{bin}/\dot{d}_{bin}=(1/2)L/\dot{L}=E/\dot{E}$,
so that \begin{equation}
\tau_{{\rm merge}}=\frac{d_{bin}}{\dot{d}_{bin}}=\frac{1}{2}\frac{m_{s}v_{bin}}{\left<F_{D}\right>}\label{eqn-tmerge}\end{equation}

In calculating the gas drag force, we must average over an orbital
period because the center of mass of the binary is moving with respect
to the background gas with relative velocity $v_{disk}$ as it orbits
the Sun. Given $m_{s}\ll m_{b}$, the relative velocity between the
binary and the disk gas is approximately the velocity at which the
large body would move through the gas on its own, given by Equations
(\ref{eqn-vrelr}) and (\ref{eqn-vrelphi}).

We now provide analytic expressions for the infall time in two different
gas-drag regimes, the regime which is linear in velocity (corresponding
to the Stokes and Epstein regimes; Section \ref{sec-linear}) and
the quadratic (ram pressure) regime (Section \ref{sec-quadratic}).
In the quadratic regime, the type of evolution depends on the ratio
$v_{bin}/v_{disk}$. In practice more complicated regimes exist (see
Section \ref{sub:Drag-laws}), which we integrate numerically for
parameters relevant to planetesimals in a protoplanetary disk in Section
\ref{sub:dependence}.

\subsubsection{Linear drag regime}

\label{sec-linear}

In the following treatment, we assume that $v_{bin}$ remains constant
over a single binary orbital period $P_{{\rm bin}}$, which is good
for $v_{bin}/\dot{v}_{bin}\gg P_{{\rm bin}}$. Note that this assumption
requires not only that $\tau_{{\rm merge}}\gg P_{{\rm bin}}/2$ but
also that $m_{s}v_{bin}/F_{D,disk}\gg P_{{\rm bin}}$, where $F_{D,disk}$
is the drag force experienced by the small body moving at relative
velocity $v_{disk}$ with respect to the gas. We address the complication
of non-circular orbits in future work.

In the linear regime, $F_{D}\propto v_{rel}$, with $v_{rel}$ equal
to the relative velocity of the small body with respect to the gas,
containing components from the binary orbit and from the overall motion
of the binary through the gas disk. Therefore $F_{D,1}\equiv F_{D}/v_{rel}$
is constant over the binary orbit. The linear regime is valid for
the Epstein and Stokes drag regimes, but the value of $F_{D,1}$ in
the two regimes differs (see Section \ref{sub:Drag-laws}). We may
now express the orbit-averaged drag force as \begin{eqnarray}
\left<F_{D}\right> & = & \frac{1}{2\pi}\int_{0}^{2\pi}F_{D}d\theta\label{eq:gas-dissipation-linear}\\
 & = & \frac{F_{D,1}}{2\pi}\int_{0}^{2\pi}(v_{bin}\sin\theta+v_{disk})d\theta=F_{D,1}v_{bin}\;,\nonumber \end{eqnarray}
 where $\theta$ is the angle of the binary in its orbit. The term
$v_{bin}\sin\theta$ is the bulk velocity component of the small planetesimal
parallel to the the direction of motion in the binary frame of reference,
so that $v_{rel}=v_{bin}\sin\theta+v_{disk}$. Over a full orbit the
contribution from $v_{disk}$ averages out and \begin{equation}
\tau_{{\rm merge}}=\frac{t_{stop}}{2}\;,\end{equation}
 with $t_{stop}$ equal to the stopping time of a single small planetesimal
in the gaseous protoplanetary disk: \[
t_{stop}=\frac{m_{s}}{F_{D,1}}=\left\{ \begin{array}{cr}
\displaystyle\left(\frac{\rho_{p}}{\rho_{g}}\right)\frac{r_s}{\bar{v}_{th}} & Epstein\\
\rule{0ex}{5ex}\displaystyle\frac{4}{9}\left(\frac{\rho_{p}}{\rho_{g}}\right)\frac{r_s^{2}}{\lambda\bar{v}_{th}} & Stokes.\end{array}\right.\]
 Recall that in the linear regime, the stopping time is independent
of the relative velocity between the planetesimal and the gas. Note
that single planetesimals with stopping times longer than an orbital
time inspiral into the star on a timescale of $\sim$$t_{stop}/\eta$. The
same processes are at work in both cases---infall into the star is
slower than binary coalescence because the gas and planetesimals orbit
the star together, reducing their relative velocities.

The timescale for coalescence is independent of $d_{bin}$, and the
total merger time for a binary is \begin{eqnarray}
T_{merge} & = & \tau_{{\rm merge}}\ln\left(\frac{d_{0}}{r_{b}}\right)\;,\end{eqnarray}
 where $d_{bin}=d_{0}$ initially, and $r_{b}$ is the final binary
separation before coalescence.

\subsubsection{Quadratic (ram pressure) regime}

\label{sec-quadratic}

We now consider the quadratic regime, for which $F_{D}\propto v_{rel}^{2}$,
appropriate for ram pressure drag. Following the same procedure as
above, but using $F_{D,2}\equiv F_{D}/v_{rel}^{2}$ with $F_{D,2}$
a constant, we get \begin{eqnarray}
\left<F_{D}\right> & = & \frac{F_{D,2}}{2\pi}\int_{0}^{2\pi}(v_{bin}\sin{\theta}+v_{disk})^{2}d\theta\nonumber \\
 & = & F_{D,2}v_{bin}^{2}\left[1+\frac{1}{2}\left(\frac{v_{disk}}{v_{bin}}\right)^{2}\right]\end{eqnarray}
 In other words, the ram pressure drag force requires an effective
relative velocity correction of $[1+0.5(v_{disk}/v_{bin})^{2}]$---in
this case the contribution from the bulk velocity drag did not average
out.

Now, \begin{eqnarray}
\tau_{{\rm merge}}=\frac{t_{stop}(v_{bin})/2}{1+0.5(v_{disk}/v_{bin})^{2}}\;,\end{eqnarray}
 where $t_{stop}(v_{bin})$ is the stopping time for $v_{rel}=v_{bin}$.
In the quadratic regime, $t_{stop}$ is not independent of $v_{rel}$,
so to make dependences clearer, we rewrite this expression as \begin{eqnarray}
\tau_{{\rm merge}} & = & \frac{m_{s}/(2F_{D,2})}{v_{bin}[1+0.5(v_{disk}/v_{bin})^{2}]}\\
 & \approx & \left\{ \begin{array}{cl}
{\displaystyle \frac{m_{s}}{2F_{D,2}v_{bin}}\;\;\;\;} & ,v_{bin}\gg v_{disk}\\
\rule{0ex}{5ex}{\displaystyle \frac{m_{s}v_{bin}}{F_{D,2}v_{disk}^{2}}} & ,v_{bin}\ll v_{disk}\end{array}\right.\nonumber \end{eqnarray}

Plugging in $F_{D,2}$ for ram pressure drag and $v_{bin}=(Gm_{b}/d_{bin})^{1/2}$,
this corresponds to \begin{eqnarray}
\tau_{{\rm merge}} & \approx & \frac{2}{0.66}\left(\frac{\rho_{p}}{\rho_{g}}\right)r_{s}\times\nonumber \\
 & \times & \left\{ \begin{array}{cl}
d_{bin}^{1/2}/\sqrt{Gm_{b}}\;\;\;\;\;\;\;\; & ,v_{bin}\gg v_{disk}\\
\rule{0ex}{5ex}2\sqrt{Gm_{b}}/(d_{bin}^{1/2}v_{disk}^{2}) & ,v_{bin}\ll v_{disk}\end{array}\right.\end{eqnarray}

Integrating, we find a total merger time of \begin{eqnarray}
T_{merge} & \approx & \frac{2}{0.33}\left(\frac{\rho_{p}}{\rho_{g}}\right)r_{s}\times\nonumber \\
 & \times & \left\{ \begin{array}{cl}
{\displaystyle \frac{\left(d_{0}^{1/2}-r_{b}^{1/2}\right)}{\sqrt{Gm_{b}}}\,\,\,\,\,\,\,\,\,\,\,\,\,\,\,\,\,} & ,v_{bin}\gg v_{disk}\\
\rule{0ex}{5ex}{\displaystyle \frac{2\sqrt{Gm_{b}}}{v_{disk}^{2}}\left(\frac{1}{r_{b}^{1/2}}-\frac{1}{d_{0}^{1/2}}\right)} & ,v_{bin}\ll v_{disk}\end{array}\right..\label{eq:T-merger-ram}\end{eqnarray}
 The merger time in the ram pressure regime depends strongly on the
ratio between the binary mutual orbital velocity and its bulk velocity
around the star. When $v_{bin}\gg v_{disk}$, the merger proceeds
most slowly when $d_{bin}$ is largest, while for $v_{bin}\ll v_{disk}$,
the final coalescence at $d_{bin}\sim r_{b}$ takes the longest time.

\subsubsection{Implications}

\label{sub:dependence}

\begin{figure}
\includegraphics[scale=0.45]{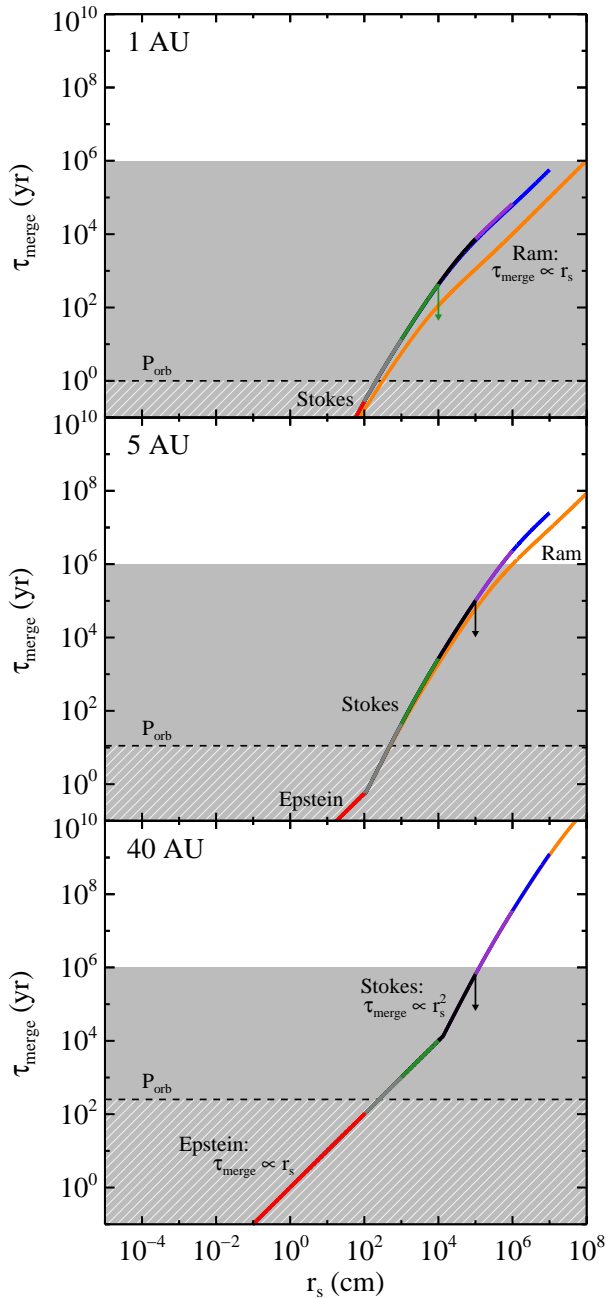}\caption{\label{fig:inspiral-parameters dependence}Merger timescales, $\tau_{\rm merge}$ for binaries located 1 (top), 5 (middle), and 40 (bottom) AU from their host stars, as a function of the radius of the smaller binary component, $r_s$.  
Binary planetesimals with a wide range of masses inspiral and likely merge on timescales shorter than the typical lifetimes of protoplanetary disks (gray region).
Colors correspond to different large planetesimal sizes, matching the colors in Figure \ref{fig:WISH-radius}.  For large planetesimals with radii $r_b \lesssim 10^3$km, the merger time is independent of $r_{b}$.  The merger time is reduced for $r_b = 10^3$km (orange, lower curve) as the ram pressure drag regime becomes important.  For $r_b < 100$m (top) or 1km (middle and bottom), our assumption that $m_{s}v_{bin}/F_{D,disk}\gg P_{{\rm bin}}$ breaks down, and our curves are upper limits.  Timescales shorter than the orbital period around the star, $P_{\rm orb}$, (hashed region) indicate that inspiral will not occur on circular orbits and a more detailed calculation is required.}

\end{figure}

As can be seen in our analytic derivation, the timescale for the inspiral
of a binary planetesimal is dependent on its environment and on the
binary properties. Figure \ref{fig:inspiral-parameters dependence}
shows the binary merger timescale as a function of small planetesimal
size for a range of big planetesimal sizes and separations from the
star. To make this figure, we calculate the integral $\left<F_{D}\right>=\int_{0}^{2\pi}F_{D}(v_{rel})d\theta$
numerically with $v_{rel}=v_{bin}\sin\theta+v_{disk}$, using the
full expression for $F_{D}$ embodied in Equation (\ref{eq:Cd}).
This, for example, allows the relevant drag law to vary as a function
of $\theta$ if appropriate. We choose either the merger timescale evaluated at $d_{bin} = min(R_{H}, R_{WS})$ or 
at $d_{bin} = r_{b}$, whichever is larger.
We maintain our assumption of circular
orbits.  This assumption is only valid for merger timescales longer than of order the orbital period of the binary around the Sun, $P_{\rm orb}$.
The assumption that $m_{s}v_{bin}/F_{D,disk}\gg P_{{\rm bin}}$ breaks down for binaries with large bodies smaller than $\sim$100m--1km in size, making our (already short) merger timescales upper limits in these cases.

We find that binary planetesimals over a wide range of masses inspiral
and likely merge in times much shorter than the typical lifetimes
of gaseous protoplanetary disks.
Binary asteroids with components having radii less than a few tens of km have been observed in the main belt
\citep[e.g.,][]{ric+06}.  Such pairs were not likely
to survive for long in a gaseous disk (see top and middle panels of Figure \ref{fig:inspiral-parameters dependence}). 
Observed
binary TNOs, however,  have radii larger than a few tens of km  \citep[e.g.,][]{nol+08b}, and could have survived for more than a Myr (bottom panel of Figure \ref{fig:inspiral-parameters dependence}).   If binary minor planets 
formed in the primordial gas disk \citep[e.g. as suggested by][]{nes+10}, those with small components are less likely to have 
survived to this day. 
Given this formation scenario, the orbital characteristics
of even the largest binary asteroids likely changed due to their evolution in
gas.  The currently observed orbital properties of
binary asteroids \citep{nao+10} are therefore unlikely to reflect only
their properties at birth; e.g. binaries born with
wide separations were likely to inspiral into more compact
configurations.  While the orbits of currently observed binary TNOs were likely unaffected,
as binary TNOs with smaller components are found, this effect will need to be considered.

\section{Other aspects of gas-planetesimal interactions}

\label{sec:other}

As we suggested above, the WISH radius can have important implications
for gas-planetesimal interactions. In this study we mainly focused
on the implications of the WISH radius for binary planetesimals. However,
this scale could be important for other processes, similar to the
role played by the Hill radius in gas-free environments. Here we only
briefly mention these issues, which will be discussed in detail (and
more quantitatively) elsewhere.

\textbf{Gas-drag induced capture and coagulation of planetesimals:
}Two unbound planetesimals may dissipate some of their kinetic energy
during an encounter (due to gas-drag), and may become bound to form
a transitional binary. Such binaries could then continue to inspiral
and finally merge due to gas drag (as discussed in the previous section).
This capture-coalescence process, which could play a role in the build
up and coagulation of planetesimals will be discussed in detail elsewhere
(Murray-Clay and Perets, in preparation; see also the settling regime
discussed by Ormel \& Klahr, 2010).

\textbf{Planetesimal erosion: }As shown above, in some regimes the
WISH radius is smaller than the big planetesimal size. This would
suggest that under these conditions wind-shearing may blow away loose
parts from the surface of single planetesimals, if they are weakly
bound to the planetesimal (e.g. pieces from aggregates held together
only by gravity could be blown from the surface of the main component
of the planetesimals). Indeed, such erosion of planetesimals was experimentally
observed for dust aggregates \citep{par+06}.

\textbf{Post impact evolution of planetesimals:} Following the collisions
of two planetesimals some of their material may be ejected from the
surface. A large fraction of this material is still gravitationally
bound to the system. However, small size particles gravitationally
bound to larger planetesimals could be blown away by the wind, if
they are ejected beyond the WISH radius. The post-impact evolution
of these particles could therefore be qualitatively different than
the corresponding non-gaseous collisional evolution, prohibiting the
smallest impact debris particles from ever accreting to the main bodies
of the planetesimals. Nevertheless, wind may also induce re-accretion
of ejecta material in some cases \citep{tei+09}, depending on the
ejecta trajectory. In addition, the short merger timescales we find
for binary planetesimals, suggest that any (WISH-stable) bound debris
around the main collision remnant would inspiral and accrete to the
main body.

\textbf{Planetesimal encounters and collisions in a gaseous environment:
}The collision rates between planetesimals vary for different regimes
of velocity and encounter distances scales, where one of the most
important scales of the problem is the Hill radius \citep[e.g.,][]{gol+04}.
The WISH radius provides an additional important parameter for planetesimal
encounters, which has to be taken into account in order to determine
the outcome of planetesimal encounters in gas. Recently (and independently)
\citet{orm+10} discussed planetesimal encounters in a gaseous environment
for some specific encounter regimes, and provided detailed calculations
for these regimes. Our study suggests an additional and complementary
understanding of these issues.

\section{Summary}

\label{sec:sum}

In this study we explored the differential gas-drag acceleration between
different size planetesimals in a gaseous environment. We defined
the wind-shearing radius as the distance at which the differential
acceleration between two close-by planetesimals is comparable to their
mutual gravitational pull. Planetesimal interactions close to or beyond
this limit would be strongly affected by wind shearing. The wind-shearing
radius has important implications for the existence and survival of
binaries. We find that binary planetesimals cannot form or survive
with separations beyond this scale, even if this separation is smaller
than the Hill radius, as they would be destabilized and sheared apart
by the head wind. WISH-stable binary planetesimal are also affected
by gas drag, and can inspiral and coalesce in times shorter than the
lifetime of the gaseous disk. The wind-shearing radius may have important
implications for planetesimal evolution, in particular planetesimal
erosion, post impact evolution of planetesimals, and planetesimal
encounters and coagulation. These effects merit further investigation.

\appendix{}

\section{Wind-shearing disruption of binary planetesimals : Parameter dependence}

\label{sub:param}

In the following we provide a more detailed discussion on the dpendence
of the WISH stability criterion for binary planetesimals on their
environment and properties.

\subsection{Gas density and relative velocity}

\begin{figure}
\begin{centering}
\includegraphics[scale=0.4]{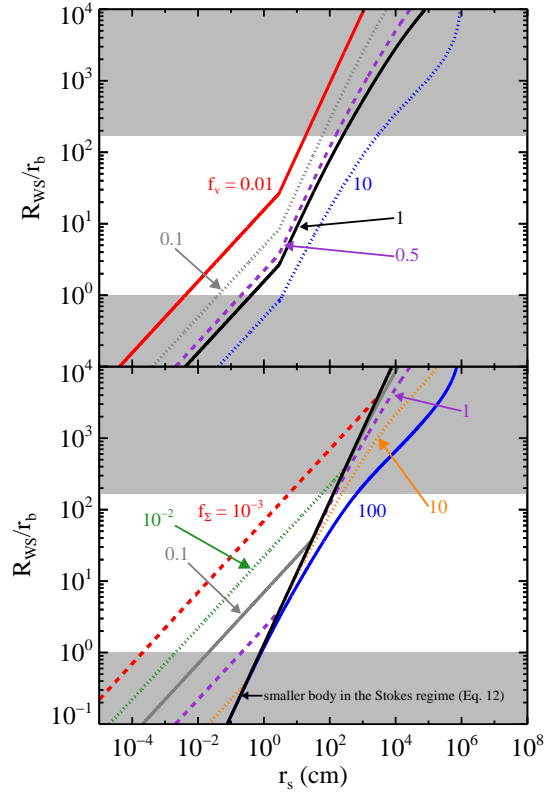}\caption{\label{fig:WISH-parameters dependence}Dependence of the the binary
WISH stability radius on the gas-density (top) and the relative velocity
between the planetesimals and the gas (bottom). These figures show
the dependence for a specific choice of the big planetesimal, $r_{b}=10$
km at $1$ AU in our fiducial disk. The density and velocity for each
line in the upper and lower panels, respectively, are $\Sigma=f_{\Sigma}\times2\times10^{3}$
g cm$^{-2}$ and $v_{rel}=f_{v}\times0.5c_{s}^{2}/v_{K}$. The dashed
lines in both panels correspond to the same parameters used for the
respective lines for the 10 km size big planetesimals shown in Figs.
\ref{fig:WISH-stability} and \ref{fig:WISH-radius}. Shaded regions
correspond to separations beyond the Hill radius (upper region) or
below the physical radius of the big planetesimals (lower region).}
\end{centering}
\end{figure}

The gaseous environment and the velocities of planetesimals with respect
to the gas likely change with time in a given disk, and vary across
planetary systems. Therefore, the WISH radius and the stability of
binary planetesimals and satellites is time and system dependent.
These issues are illustrated in Figure \ref{fig:WISH-parameters dependence},
which shows how the WISH radius varies as a function of the relative
velocity and gas density. The figure shows the WISH stability radius
of an $r_{b}=10$ km planetesimal at $1$ AU, for various choices
of the relative planetesimal-gas velocity (upper panel) and the gas
density (lower panel). Note that for small planetesimals in the Stokes
regime the WISH radius becomes independent of the gas-density, as
can be seen in Eq. (\ref{eq:r_WS_all}), and the various lines in
Figure \ref{fig:WISH-parameters dependence} converge. We do not show
the dependence of $R_{WS}$ on temperature since it is weak (at most
$R_{WS}\propto c_{s}^{-1/2}\propto T^{-1/4}$ ; see Eq. \ref{eq:r_WS_all}).

As might be expected, the general trend of the WISH radius is to be
smaller for higher gas densities and/or higher planetesimal velocities
relative to the gas, i.e. WISH becomes more pronounced with stronger
gas drag. Also, as mentioned before, very large comparable size planetesimals
are hardly affected by gas-drag and the WISH radius becomes larger
than their Hill radius. Taken together, the WISH will be more important
for smaller binary planetesimals, during earlier stages of their growth/evolution,
in a more gas-rich environment; WISH gradually becomes negligible
at later stages.

\bibliographystyle{apj} 
%\bibliographystyle{apj}
%\bibliography{planet-formation}

\end{document}